# Case Study on Cloud Based Library Software as a Service: Evaluating EZproxy

[1]**Robert Iles,** [2] **Emre Erturk**

[1] Electronic Services Librarian, Eastern Institute of Technology, New Zealand

[2]Senior Lecturer, School of Computing, Eastern Institute of Technology, New Zealand

E-mail:  [1]riles@eit.ac.nz, [2]eerturk@eit.ac.nz

## ABSTRACT

There is a growing relationship between academic libraries and cloud computing. Therefore, understanding the beginnings and the current use of cloud base services in libraries is important. This will help understand the factors that libraries should consider in the future. The purpose of this paper is to better understand the future implementation of the cloud based software in academic settings. Using cloud based, web based, and other remote services may bring both advantages and disadvantages, some of which this paper will bring out. First, a brief literature review of the academic literature, and a review of available general-purpose cloud-based library products are conducted. Next, a real-life scenario for a mid-sized New Zealand institution of higher education is evaluated. This case involves moving from a locally hosted version of EZproxy to a cloud based version with support from the vendor. As this information system decision is an important one, this paper makes a contribution to the available literature and can be informative for librarians. In conclusion, academic libraries will gradually involve more pervasive use of cloud based systems. The examples of important factors to be considered in future decisions include timing and staffing.

**Keywords:** *Cloud Computing, SaaS, Library Systems*

## 1. INTRODUCTION

Cloud computing is a technology delivery model that provides ubiquitous and on-demand applications and services, by utilizing online and remote resources that can be provisioned with minimal management or service provider interaction [13]. One example of this is Software as a Service (SaaS), in which the software is hosted by a service provider and accessed by the client online, for example, through a web browser. The paper first discusses the early history of the access to cloud hosted or based environments along with the current state of affairs. Next, it moves onto looking at how this may expand in the future. The first part of this research is based on the academic and professional literature along with a survey of the library software vendors and their websites. The second part of this research involves a brief case study that pertains to a mid-sized New Zealand institution of higher education. The main academic library at the institution uses EZproxy, which comes in two versions. One version is cloud hosted and operated by the vendor, while the other version is locally installed and maintained. It is important to look at the advantages and disadvantages of each version. This analysis leads to implications and decisions for this institution as well as others in a similar situation that are contemplating how much of their library services need to be cloud based.

## 2. THE SPREAD OF CLOUD BASED SOFTWARE IN LIBRARIES

From the beginning, computer technology and networking have brought an abundance of new electronic resources to schools and libraries, and have made a significant impact on education, training, and human resources in general [5]. Academic libraries' association with cloud computing can be seen as the nascent relationship between the hosted services with commercial library vendors. These vendors started to make resources available over the web in the 1990's. These services included full text library databases mostly comprised of journal materials. First, they were only available on campus via IP authentication but, with the advent of tools such as EZproxy, these resources became available off campus. This tied in well with the growth in ownership of personal computers by students and their desire to study off-campus. The growing usefulness of remotely hosted resources laid the ground work for vendor-provided cloud-based library solutions as the next logical step.

As a result, library vendors have changed their services to cloud based ones, whereby the library buys a subscription. Mostly these have been full text journal or eBook services. However, this is starting to be expanded into library catalogues, discovery tools, and multimedia [14]. Discovery tools such as ExLibris's Primo can be purchased either as a cloud service or locally hosted. The main advantage of the cloud version is the cost saving in terms of computer hardware. The cloud version comes in two varieties. The first one is just hosted and the client does maintenance, customization, and installs the software updates. The other option is fully maintained and updated by ExLibris but is a lot less customizable. Online Computer Library Center (OCLC) has brought out a multi-tenant Library System called "World Share" which is cloud based. This brings together a few of the already successful tools for libraries such as "World Share Interlibrary Loan" and wrapped them into a unified environment based around the features of a traditional library system such as cataloguing, acquisitions, and Online Public Access Catalog [1]. Overall, cloud based Software as a Service (SaaS) has reduced the need for local library IT support and hardware procurement [2].

Furthermore, the advent of Bring Your Own Device (BYOD) in academic libraries has brought in new uses for cloud computing. The provision of robust wireless networks in academic libraries has been essential to help provide cloud based tools such as Google Docs, Dropbox, as well as access to library services that are cloud based [11]. There is also a need for current library services to be mobile friendly for users of library tools such as the Library Management System (LMS) and Discovery tools [4].

| Company | Name of Product | Type of Product |
|---|---|---|
| ExLibris | Aleph | Library Management System |
| ExLibris | Alma | Library Management System |
| ExLibris | Primo | Discovery Layer |
| OCLC | WorldShare | LMS and other products |
| ProQuest | Summons | Discovery Layer |
| ProQuest | Intota 2 | Library Management System |
| III | Sierra | Library Management System |
| LibLime | Koha | Open Source LMS |
| Sirsidynix | BLUEcloud LSP | Library Management System |
| EBSCO | EDS | Discovery Layer |

**Table 1.** Sample Library System Vendors with Cloud Services

Libraries have also started experimenting with the use of cloud computing on their own rather than just through obtaining services from vendors. For example, local resource collections (such as photo libraries) for individual institutions have been set up online [7]. On the other hand, the use of cloud computing in academic libraries has been somewhat affected by the perceived risks of cloud computing and due to lack of end user understanding. The main perceived risk seems to be about trust in relation to data sovereignty [16]. This risk is overcome partly by hybrid and private cloud environments [3] but this in turn lessens the financial benefits to be gained from a purely public cloud usage. Lack of understanding about the role of cloud computing in libraries is also decreasing with an array of literature published on this subject [9].

## 3. THE FUTURE USE OF CLOUD COMPUTING IN LIBRARIES

Many academic libraries are facing cost cutting measures. At the same time, there is an ongoing desire to improve services for users to access library information. The pressure from users to embrace new technologies causes librarians to look more closely into cloud computing [15]. Other key drivers are the saving of money on hardware procurement (which is an upfront cost) and the constant growth of rich electronic resources. SaaS has been widely accepted by academic libraries, and tends to be vendor driven. But the strengthening of wireless networks and BYOD has brought about reliance on online storage and library clients demanding the new cloud based services. For example, Google Drive and Office 365 provide storage for students' work, as well as the actual software for creating content and collaboration. As teaching and learning pedagogy become more collaborative, these tools become more important. Many academic libraries are teaching information literacy and how to use different cloud based tools, including Google Scholar [10]. With data centers increasing and becoming more localized, cloud based library services will grow. However, privacy issues around user information and issues relating to long haul internet connections are also important. New Zealand can benefit from data centers located domestically or at least in Australia in order to increase the speed and alleviate the security factors, which have until now detracted academic libraries from using the cloud. Fast research networks such as REANNZ/KAREN that are available to academic libraries can help although bandwidth may be a limiting factor to some degree, especially for users off campus. When Netflix was introduced in Australia, it affected many Australian internet users and their speed. For some users, limited bandwidth can make it difficult to multitask with other cloud software [8]. As cloud computing becomes main stream, the advantages will also become more apparent. Trust and understanding cloud computing will also increase its usage, especially as the benefits of community, private, public and hybrid cloud models become apparent.

Some cloud based library technologies are starting to mature, e.g. cloud based discovery tools. This will enhance their uptake within academic libraries [14]. Another important turning point in using new technology takes place when academic libraries have to update current systems or implement additional services. The future will also bring about a greater use of the already implemented aspects; the cloud will become more pervasive within academic libraries. Infrastructure as a Service (IaaS) type repositories such as D-Space will expand, particularly advantages (such as scalability) of sharing and combining resources become apparent, compared to individual and locally hosted. Due to greater storage and scalability, cloud computing can provide better business intelligence through the analysis of the so-called big data [6]. This could be an advantage for an academic library in helping to understand how its services are used. In addition, if clear economic advantages are forecast, they will add to the driving force behind adopting cloud services [12].

The relationship between the cloud and academic libraries has mostly been driven by library software vendors. Nevertheless, academic libraries have also been ready to take on new services and methods of delivery. Some cloud usage within academic libraries is driven by the end user, for example the proliferation of BYOD along with the related cloud based applications. Uptake of the cloud by academic libraries is increasing as the quality of the services improves and security concerns are addressed, along with a better internet infrastructure and understanding of cloud computing.

## 4. CASE STUDY ON A CLOUD BASED LIBRARY TOOL

EZproxy is a software application that provides off campus access to library provides databases such as those from ProQuest and EBSCO along with specialized local databases such as Standards New Zealand. Before the implementation of EZproxy, databases were only available on campus and were authenticated through the campus IP address. Database providers would give access only to users coming from an IP address that had been lodged with them. More flexibility on access from outside of the campus was expected as online learning environments such as Blackboard and Moodle grew. EZproxy filled this gap by facilitating reliable access to library's subscription databases. Figure 1 is an outline of how EZproxy works.

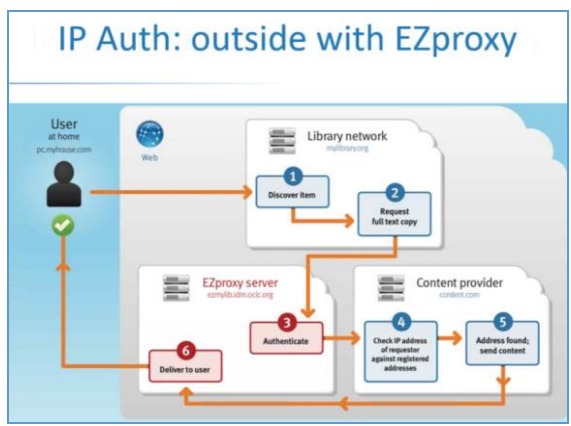

**Figure 1**: IP Authorization off-campus with EZproxy (from OCLC, 2015)

There is a JavaScript based tool that library staff can use to create EZproxy links for resources in the library databases. This small application is simple and easy to use, and provides a way to create library links on Moodle course sites. See Figure 2 for a screenshot.

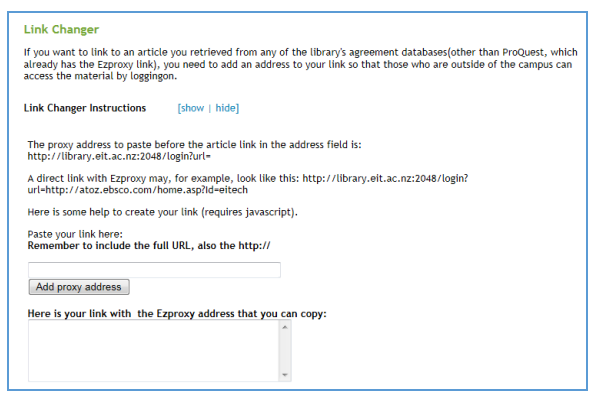

**Figure 2**: EZproxy Link Changer

With the advent of cloud computing, it was inevitable that EZproxy (see Figure 3 below for a screenshot) would be installed and run in a hosted cloud environment. This would also remove some of the overheads that the locally hosted EZproxy software places on a library. This case study looks at the pros and cons of moving to the cloud based version of EZproxy. It also recommends the most advantageous timeframe and conditions for such a migration.

The main academic library at this institution introduced EZproxy from Useful Utilities in 2006. The competitive selection process at the time also considered Novell's Netware iChain but EZproxy's strong record for providing off campus access to library databases was seen as an advantage. The EZproxy software was then installed on the existing library server. A licence was purchased for US$ S60 from EZproxy. IT staff with help from Chris Zagar (the original founder of EZproxy and Useful Utilities) got it working using Lightweight Directory Access Protocol (LDAP). The network port name was used for the set up rather than the host name. The Electronic Services Librarian tested the configurations for the databases that the library had subscribed to. Configuration and maintenance of the software was supported not only by Useful Utilities but also assistance through the EZproxy Listserv. The active EZproxy List Serve is an extremely useful place to get help with EZproxy. Browsing the archives will often provide database configurations that have been successfully tested elsewhere and can be reused.

Useful Utilities was sold to OCLC in January 2008. The move to OCLC was advantageous as it brought with it the resources of a large organization. However, the licensing fee went up. This price increase did not take place immediately but after a hiatus of about five years. Over the years, the number of library databases at this institution has increased.

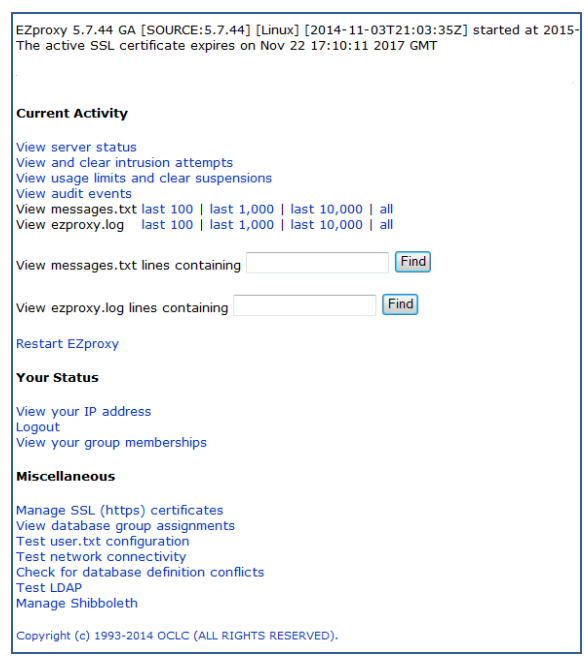

**Figure 3**: Screenshot of EZproxy Web Interface

The upgrades to EZproxy took place once a year with little overhead in terms of staff time and costs. In 2014, the library server moved from Windows to Red Hat Linux to accommodate the Voyager library system. EZproxy was upgraded and moved to the new platform at the same time. The current annual licensing fee cost to the institution is approximately US$ 400. The most recent version for the Linux platform (as of this writing) is 6.0.8.

If EZproxy is locally hosted, the security is the responsibility of the organisation; but with the cloud version, OCLC presumes responsibility. However, since the individual institution signs the agreement with the library database provider, OCLC does not have to deal directly with the database provider if a breach occurs. In that situation, the institution may also risk having their database access terminated. Furthermore, user authentication in the cloud version will occur between the institution's AD/LDAP server and OCLC's server. This may be prone to a 'man-in-the-middle' hacking attack. Backups of the EZproxy data are done by the institution's IT department. The EZproxy files at this institution have a total size of about 500 MB, consisting mainly of logs. This is relatively small, and not a drain on internal IT resources and processes.

## 5. FURTHER DISCUSSION OF THE CLOUD BASED SOLUTION

For the cloud based version of EZproxy, the vendor company conducts the monitoring, maintenance, software updates, and backups and recovery. The time frame for changing over to a cloud based version of EZproxy is one of the crucial issues. The library already has a server for hosting the main library system (Voyager). Therefore, while this in-house server is in use and available, hosting EZproxy locally is not a major issue. When the Voyager library system is retired in the future, the next generation of library systems will almost certainly be cloud based, without any need for a local server. This will probably be when EZproxy will also be moved to the cloud. Another aspect that needs to be tested is the speed of access to the Australian cloud hosting of EZproxy, given that most of our students are in New Zealand while most of the databases are hosted in the United States of America. Routing through Australia (rather than directly through Hawaii) would slightly increase the distances that the data packets travel.

Current cost for a locally hosted EZproxy license is currently about NZ$ 650 for the institution. For cloud hosting, this would jump to anywhere between $1500 and $4000 NZD (depending on the institution's student enrolments). Therefore, there would be an increase in cost. There is also another implementation fee, which (although fairly minor) would also need to be added. The institution would therefore have to be certain that other advantages would stack up in order to make the cloud based version of EZproxy software a viable option.

New database deployment may be an issue with the cloud based EZproxy with a quoted time of two days for a new database deployment. Deployments do not happen regularly; however, the faster each deployment, the better it is for the end users. Currently, it is possible for internal staff to deploy a new database within two hours, with some more time for off campus testing. In an email message on May 28, 2015, Angus Cook from OCLC has informed the institution that a new administration module will be released, which may facilitate future deployments

## 5. CONCLUSIONS

The cloud based applications that are used by academic library users are not limited to resource databases, but also other applications that students regularly use in conjunction, while reading references and writing assignments. Therefore libraries need to leverage the increasing end user information literacy levels. This can be done by better interfacing library systems with popular cloud based tools such as Dropbox, Google Docs and Office 365.

EZproxy, as a core library application, offers a cloud hosted software as a service version, which seems to be the best option in the long term. This would diminish need for the library to have its own server. Timing of the move will depend on other things such as the institution's happiness with the level of security and making a formal decision to retire Voyager. Currently, internally supporting EZproxy is not a problem with the staff levels and skills at hand. However, if staff turnover or reductions were to take place, they would accelerate the move to the cloud based version of EZproxy software. Financial and human resource related issues pertain to many libraries around the world. Therefore, they will come across these similar factors as this institution while deciding between local management of library systems versus remotely hosted or outsourced services that may also be shared with other institutions.

## AUTHOR PROFILES

**Emre Erturk** received his Ph.D. from the University of Oklahoma in 2007. He has later taught with the University of Maryland (USA) in the areas of business, economics, and IT. Currently, he is an IT Lecturer at the Eastern Institute of Technology in New Zealand. He has been involved in many conference presentations and publications around the world.

**Robert Iles** has been working as a Librarian at the Eastern Institute of Technology since 2006. Previously, he has studied with Massey University and University of Waikato in New Zealand. He is currently also a post-graduate student at the Eastern Institute of Technology.